# Reducing Graphene Device Variability with Yttrium Sacrificial Layers

Ning C. Wang[1], Enrique A. Carrion[2], Maryann C. Tung[1], Eric Pop[1,3,a]

*[1]Department of Electrical Engineering, Stanford University, Stanford, CA 94305, USA*

*[2]Department of Electrical and Computer Engineering & Micro and Nanotechnology Lab, University of Illinois at Urbana-Champaign, Urbana, IL 61801, USA*

*[3]Department of Materials Science & Engineering, Stanford University, Stanford, CA 94305, USA*

Graphene technology has made great strides since the material was isolated more than a decade ago. However, despite improvements in growth quality and numerous "hero" devices, challenges of uniformity remain, restricting large-scale development of graphene-based technologies. Here we investigate and reduce the variability of graphene transistors by studying the effects of contact metals (with and without Ti layer), resist, and yttrium (Y) sacrificial layers during the fabrication of hundreds of devices. We find that with optical photolithography, residual resist and process contamination is unavoidable, ultimately limiting device performance and yield. However, using Y sacrificial layers to isolate the graphene from processing conditions improves the yield (from 73% to 97%), average device performance (three-fold increase of mobility, 58% lower contact resistance), and the device-to-device variability (standard deviation of Dirac voltage reduced by 20%). In contrast to other sacrificial layer techniques, removal of the Y sacrificial layer with HCl does not harm surrounding materials, simplifying large-scale graphene fabrication.

[a] Author to whom correspondence should be addressed. Electronic mail: epop@stanford.edu

Since the first experimental demonstration of monolayer graphene in 2004,[1] academic and industrial research labs have extensively explored applications that leverage the unique electrical, mechanical, and thermal properties of this material. To date, these efforts have yielded promising results for high-sensitivity chemical sensors,[2] transparent and flexible electronics,[3] and analog circuits.[4] Rapid advances in large-scale production of graphene via chemical vapor deposition (CVD) have also accelerated the development and mass-production of graphene towards practical applications.[5]

While unique physical properties motivate research in graphene electronics, device yield and spatial variability will ultimately dictate industrial impact. To this end, many works have focused on improving quality and cost of large-scale CVD graphene growth and transfers,[6,7] but fewer have attempted to quantify and reduce the detrimental effects of subsequent device fabrication, which may lead to inconsistencies in reported results. Optical photolithography—critical for large-scale fabrication—is particularly problematic as photoresist (PR) residues on the graphene surface negatively impact device performance.[8] Additionally, post-fabrication residue removal methods, such as non-selective physical etches with $CO_2$,[9] are difficult to reproduce in a controlled manner and may damage graphene by inducing tears and wrinkles.

In contrast to post-fabrication cleaning methods, modifying and optimizing a process flow to protect graphene from its interaction with PR is a more suitable approach. The semiconductor industry has employed native $SiO_2$ as a sacrificial protection layer for Si during integrated circuit fabrication. However, since graphene has no equivalent native oxide, thin aluminum (Al) and titanium (Ti) layers have been introduced as sacrificial layers.[10,11,12] Subsequent removal is nontrivial, and can affect graphene and surrounding materials (i.e. substrates, insulators, PR) during fabrication. For example, Al removal via AZ422 developer[12] lengthens resist development times, potentially leading to overdevelopment and reduced yield of small features. For Ti removal,[10] hydrofluoric (HF) acid was utilized. However, HF can etch underneath the graphene and delaminate devices from the substrate, lowering device yield on oxide substrates such as $SiO_2$.

In this work, we introduce the use of yttrium (Y) sacrificial layers to protect graphene. Y is an ideal material for this purpose, as it (1) readily forms a sub-stoichiometric oxide that does not degrade electrical transport[13] and (2) is etched in hydrochloric acid (HCl), which is safe for both $SiO_2$ and PR. The latter point greatly simplifies fabrication, resulting in higher device yield and differentiates the Y-sacrificial method from existing alternatives. We fabricate and measure hundreds of devices, employing transport models and materials analysis to physically quantify the cause of fabrication-induced degradation. We then introduce the use of Y-sacrificial layers to alleviate degradation, reduce variability, increase performance (three-fold intrinsic mobility improvement and 58% decrease in contact resistance), and improve our device yield up to 97%.

Figure 1 outlines the process flow used, including implementation of the Y-sacrificial layer. Complete fabrication details are given in the Supplementary Material, and we outline here materials characterization relevant to Y-sacrificial layer processing. Monolayer graphene is first grown on Cu foils by chemical vapor deposition at 1000 °C, using isopropyl alcohol (IPA) or $CH_4$ carbon source (additional details are given in the Supplementary Material). The graphene is then transferred onto 90 nm $SiO_2$ on p$^+$ Si ($< 0.005$ $\Omega\cdot$cm) substrates with a poly(methyl methacrylate) (PMMA) support scaffold[14] [Fig. 1(a)], applying a modified RCA cleaning process[7] to minimize wrinkles and impurities (i.e. Fe, Cl, Cu). Immediately after transfer and prior to device fabrication, the samples are coated with ~5 nm of Y via electron-beam evaporation in ~$10^{-7}$ Torr base pressure to isolate and protect the underlying graphene [Fig. 1(b)]. The Y-coated samples are not metallic, indicating that the Y film is oxidized upon exposure to air after deposition.

To characterize the effects of process conditions on device quality, we fabricate back-gated graphene field-effect transistors (GFETs) using ultraviolet (UV) photolithography, as shown in Fig. 1(c-h). During Y deposition, a glass slide raised by silicon spacers shields half of each sample to yield control regions where the graphene and subsequent devices are fully exposed to process conditions. This approach leads to two types of devices, as shown in Fig. 1(e), "bare" GFETs with potential PR residue and/or processing damage in the channel and under the contacts, and "Y protected" devices.

We also repeat GFET fabrication (with and without Y protection) using pure palladium (Pd) contacts instead of Ti/Pd contact stacks (where Ti serves as the adhesion layer) and polymethylglutarimide/Shipley 3612 (PMGI/SPR3612) instead of lift-off layer 2000/Shipley 3612 (LOL2000/SPR3612) bilayer PR stacks. By eliminating the Ti adhesion layer which can oxidize, the Pd contacted devices should exhibit cleaner, more ideal contact interfaces and lower contact resistance.[15] Derived from the same chemistry, PMGI and LOL2000 differ only in the fact that LOL2000 contains additional contrast-enhancing dye. Thus, use of PMGI may lower the amount of residual PR. These process splits allow us to compare improvements from complete graphene protection with Y-sacrificial layers vs. other common process variations.

The Y sacrificial layer is etched in 10:1 DI:HCl after resist exposure and development to expose the underlying graphene when necessary, such as prior to contact metallization [Fig. 1(c)], channel definition [Fig. 1(f)], and final device definition [Fig. 1(h)]. Contact metals for all devices remained intact despite the total of three HCl baths necessary for back-gated GFET fabrication, with no optically visible delamination or lateral removal of metal. AFM measurements show ~1 to 1.5 nm contact metal reduction after a single 15 second bath (Supplementary Figure S3), yielding an HCl etch selectivity ~3:1 of Y to the contact metal (here 1.5 nm Ti capped with 40 nm Pd). Processing with sacrificial Y should not affect device yield for contact metals of typical thickness (> 40 nm) since Y layers are thin (< 5 nm) and etch times are short (at most 60 seconds).

Auger spectroscopy results shown in Fig. 2(b) and Supplementary Fig. S2 confirm Y removal in desired regions, and the ~1736 eV peak present only over the channel is indicative of a protective, partially oxidized Y layer.[16] Raman spectroscopy of graphene after Y removal also indicates that no graphene damage is induced (Supplementary Figure S4). To assess sample cleanliness, we conduct atomic force microscopy (AFM) of graphene in the "open" contact regions [Fig. 2(c-d)] after PR exposure, development and Y etching, but prior to metal deposition. Root mean square (RMS) surface roughness decreases by a factor of two (from 1.29 to 0.69 nm) with the use of a Y-sacrificial layer, consistent with a cleaner surface.

Figure 3 displays a summary of electrical data for the subset of 44 bare and 56 Y-protected devices fabricated using Ti/Pd contacts and LOL2000 resist, and the arrows display the changes between them. Electrical measurements are taken in vacuum ($T = 294$ K, $P \approx 10^{-5}$ Torr) following an *in situ* vacuum anneal at 200 °C for 1 hour, which also reverses remnant HCl-induced doping (Supplementary Figures S5 and S6). $I_D$ vs. $V_{BG}$ measurements (at $V_{DS} = 50$ mV) as in Fig. 3(a) are then fit to our charge transport model[17,18] using a least-squares method to extract contact resistance ($R_C$), carrier puddle density ($n^*$), and intrinsic mobility ($\mu$). The total device resistance is $R = (L/W)R_S + 2R_C + R_{series}$, where $R_S = [q\mu(n+p)]^{-1}$ is the graphene sheet resistance, $q$ is the elementary charge, and ($n+p$) is the total carrier concentration which depends on $V_{BG}$ (reaching a minimum at $n + p = 2n_0$ when $V_{BG} = V_0$, the charge neutrality point).[17,18] We note that $n_0 = [(n^*/2)^2 + n_{th}^2]^{1/2}$, where $n^*$ is the carrier puddle density generated by ionized impurities, and $n_{th}$ are the thermally-generated carriers.[17,18] Thus, at a given temperature, lower $n_0$ means lower graphene impurity density. $R_C$ follows a transfer-length method (TLM) model,[17] and both it and the mobility are listed at a carrier density of $5 \times 10^{12}$ cm$^{-2}$ in the remainder of this work. $R_{series} = 12$ Ω is our total parasitic pad, lead, and cable resistance.

Figure 3(a) shows much improved transport in Y-protected samples due to cleaner metal-graphene contact interfaces. Devices exhibit electron-hole asymmetry due to use of large metal work-function contacts, which increases contact resistance for electrons and leads to lower current at posi-tive $V_{\text{BG}} - V_0$.[19] Figure 3(b) reveals that the Y-sacrificial process decreases the estimated puddle carrier density $n^*$ from $4.6 \times 10^{11}$ to $2.2 \times 10^{11}$ cm$^{-2}$, which is indicative of a cleaner graphene surface. In fact, the puddle carrier density of the Y-protected CVD graphene samples is comparable to that of previous studies on exfoliated graphene on SiO$_2$.[17,18] This is an important device metric, as it sets a limit on the minimum GFET current, and therefore on the maximum achievable on/off current ratio. In the ideal case of impurity-free graphene, the minimum carrier density would be given by thermally-generated carriers, $2n_0 = 2n_{\text{th}} \approx 1.6 \times 10^{11}$ cm$^{-2}$ at 300 K.[18] (Thus, an on/off ratio >100 could be achieved in ultra-clean GFETs at room temperature, depending on contact resistance and assuming $>10^{13}$ cm$^{-2}$ maximum carrier density achieved via strong gating. To date, the highest GFET on/off ratios we are aware of are ~24 at 300 K and ~35 at 250 K, for graphene samples encapsulated by h-BN.[20])

In addition, Figs. 3(c-d) show that average $R_{\text{C}}$ and $\mu$ improve by factors of ~2.5x and ~3x respectively ($R_{\text{C}}$ decreases from 4.67 to 1.92 k$\Omega$ $\mu$m while $\mu$ increases from ~1200 to ~3100 cm$^2$V$^{-1}$s$^{-1}$). While these values are for holes, similar improvements in $R_{\text{C}}$ and $\mu$ are also observed for electron transport (Supplementary Figure S7). Table I summarizes the electrical properties and yield of 227 measured devices using atmospheric-pressure CVD graphene, which includes those fabricated with pure-Pd contacts and using PMGI lift-off resist as well. We use a conservative definition of yield, i.e. the percentage of devices with a well-defined, single charge neutrality point (CNP) $V_0$ and on/off ratio > 3, where on-current is measured at maximum negative gate bias ($V_{\text{BG}}$ = -30 V) and off-current is measured at the CNP. This expresses the yield in terms of electrically "well-behaved" devices, which is a better indicator of useful devices than simply the proportion of electrically active devices.

When comparing the device splits processed *without* the Y-protective layer (Table 1, LOL2000 resist and Ti/Pd contacts vs. PMGI resist and Pd contacts), only yield improves as the number of devices with single CNP increases for the PMGI process. Since multiple CNPs signify the presence of charged interface traps,[21] the improvement in yield is indicative of cleaner graphene interfaces with use of PMGI vs. LOL2000, where the contrast-enhancing dye may leave surface residue. However, despite a more ideal metal-graphene interface with Pd-only contacts, the lack of average device performance improvement suggests another mechanism is suppressing device performance. The hypothesis is supported by analyzing the Y-sacrificial layer devices, which exhibit both mobility, impurity density, *and* contact resistance improvement over "bare" devices, regardless of metal and/or resist used (Table 1). This indicates that process-induced contamination ultimately limits the performance of graphene devices, and attempts to reduce residue are insufficient. In other words, graphene must be *fully* shielded from fabrication conditions in order to obtain optimal device performance.

To demonstrate the wide applicability of our approach, we fabricate and measure additional devices using graphene from various academic and commercial sources (our atmospheric vs. low-pressure CVD growth vs. Graphene Supermarket, see details in Supplement). The extracted average values of mobility, contact resistance and carrier puddle density (Supplementary Table S1) show a clear improvement for all GFETs fabricated with Y-sacrificial layers, underscoring the effectiveness of the technique. To further quantify the variability reduction, we extract the CNP, also known as the Dirac voltage $V_0$, for all measured devices (Supplementary Figure S8). The average $V_0$ is reduced from 1.22 to 0.89 V and the standard deviation decreased from 3.19 to 2.54 V with the Y-sacrificial process. Given that our test devices are using 90 nm $SiO_2$ as the back-gate dielectric, the equivalent $V_0$ for 1 nm thin $SiO_2$ would be 0.01 ± 0.028 V, very close to the true charge neutrality, signifying that the Y-sacrificial layer technique does not induce additional doping.

The impact of process-induced contamination on large-scale GFET fabrication is also evident from the measured on/off ratios of the subset of atmospheric-pressure derived graphene devices, as shown in Fig. 4(a). Regardless of PR stack or contact metal, bare devices (without a Y-sacrificial layer) have higher $R_C$ due to process-induced contamination, yielding lower on/off ratio at shorter channel lengths as $R_C$ begins to dominate. This problem is alleviated with use of Y-sacrificial layers, which enable cleaner contacts and preserve the on/off ratio down to the shortest channel lengths used in this study ($L = 3$ μm). In Fig. 4(b) we also quantify a simple analog device metric, the maximum ratio of transconductance to current max($g_m/I_D$), extracted from $I_D$ vs. $V_{BG}$ sweeps. The Y-sacrificial layers improve average max($g_m/I_D$) by ~60%, reflecting an improvement in channel quality as the max($g_m/I_D$) is obtained at biases near $V_0$, where the channel dominates device transport. Since the oxide is the same for both types of samples, the improvement is solely from the Y-sacrificial layer technique, reinforcing the notion that graphene must be protected from PR during processing to maximize device performance.

In summary, we introduced a simple and practical method to improve graphene device quality through the use of Y-protective layers during processing. By electrically characterizing hundreds of graphene devices fabricated under various conditions, we identify through physical, quantitative analysis that PR residue restricts device variability and performance, hampering large-scale analysis. Unlike existing methods, the Y-sacrificial technique protects the graphene without affecting surrounding materials, leading to increased device performance, reduced variability, and increased yield (here up to 97%). Although demonstrated for GFETs on $SiO_2$, the Y-sacrificial technique is applicable to other HCl-compatible two-dimensional materials and substrates, where process contamination could be a concern.

See supplementary material for further details and analysis of yttrium sacrificial layer process as well as additional electrical data.

This work was supported in part by Systems on Nanoscale Information Fabrics (SONIC), one of six SRC STARnet Centers sponsored by MARCO and DARPA, by the Air Force Office of Scientific Research (AFOSR) grant FA9550-14-1-0251, in part by the National Science Foundation (NSF) EFRI 2-DARE grant 1542883, and by the Stanford SystemX Alliance. We thank Ling Li and Prof.

H.-S.P. Wong for providing the CVD graphene samples from Graphene Supermarket. We thank Sam Vaziri for helpful comments during the manuscript preparation process.

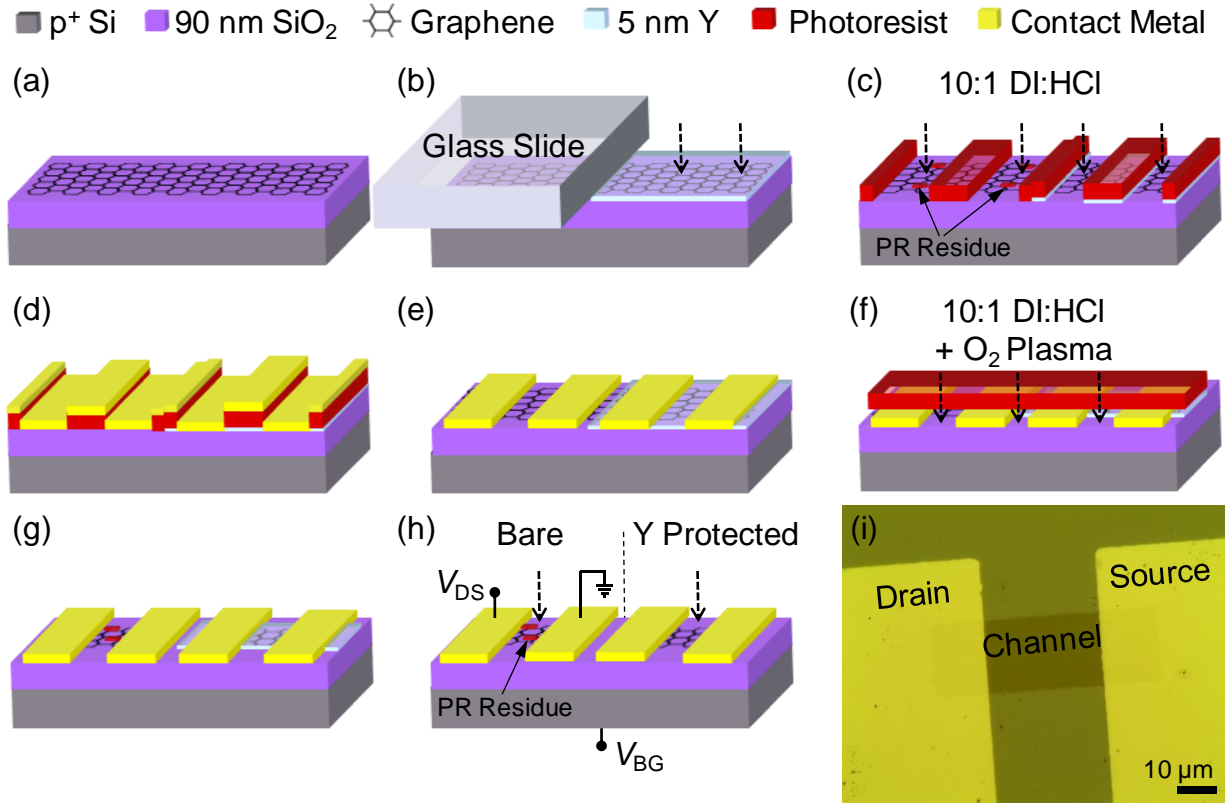

**FIG. 1**. Side-by-side fabrication steps of control ("bare") graphene devices and of those protected with the Y sacrificial layer. (a) CVD-grown graphene is transferred[14] onto 90 nm SiO$_2$ on Si after modified RCA clean.[7] (b) ~5 nm Y deposition via electron beam evaporation. Half of sample is covered for the control devices. (c) Contact metal photolithography and subsequent Y removal in dilute 10:1 DI:HCl. Some PR residues remain on control devices. (d) Deposition of contact metal by electron beam evaporation (here Ti/Pd or pure Pd). (e) Devices after metal liftoff. (f) Channel photolithography, Y removal, and graphene removal (O$_2$ plasma). (g) Defined devices after PR removal. (h) Final devices (channel lengths $L$ = 3−10 μm and widths $W$ = 3−20 μm) after subsequent channel lithography and final HCl clean. (f) Optical image of single device which underwent Y processing.

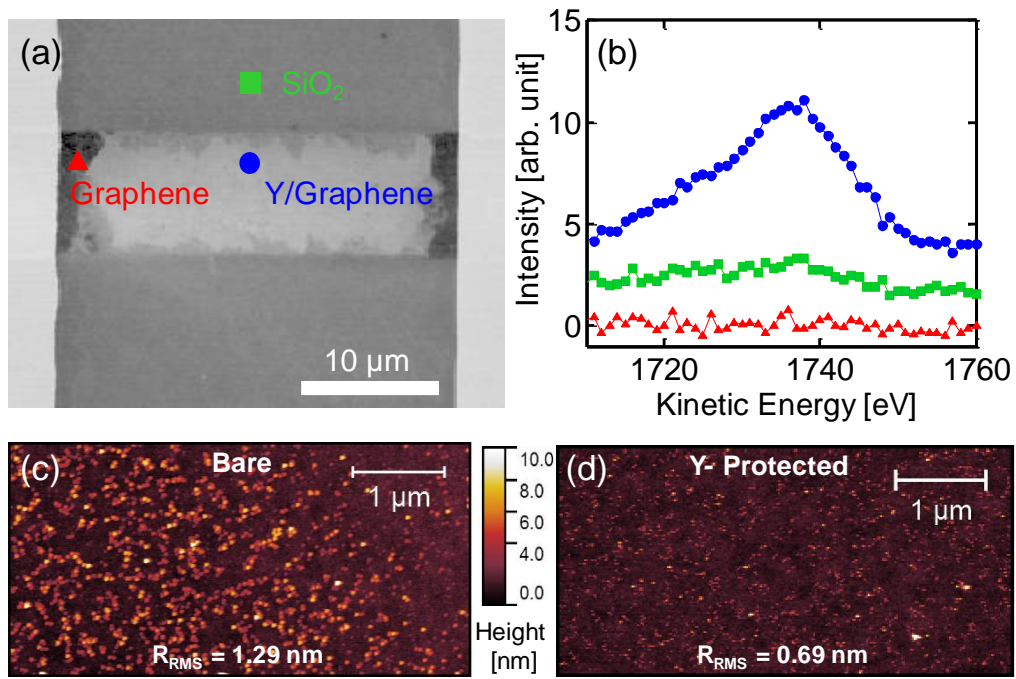

**FIG. 2.** (a) Scanning electron micrograph (SEM) of device after channel definition and before final HCl bath. Symbols correspond to acquisition points for Auger spectroscopy shown in (b). The ~1736 eV peak present only over the channel is lower than known peaks associated with elemental Y and is indicative of partially oxidized Y.[16] AFM surface comparison of (c) bare (control) and (d) Y-protected graphene devices at contact regions after photolithography and prior to metallization. Using Y-sacrificial layers decreases RMS roughness from 1.29 to 0.69 nm.

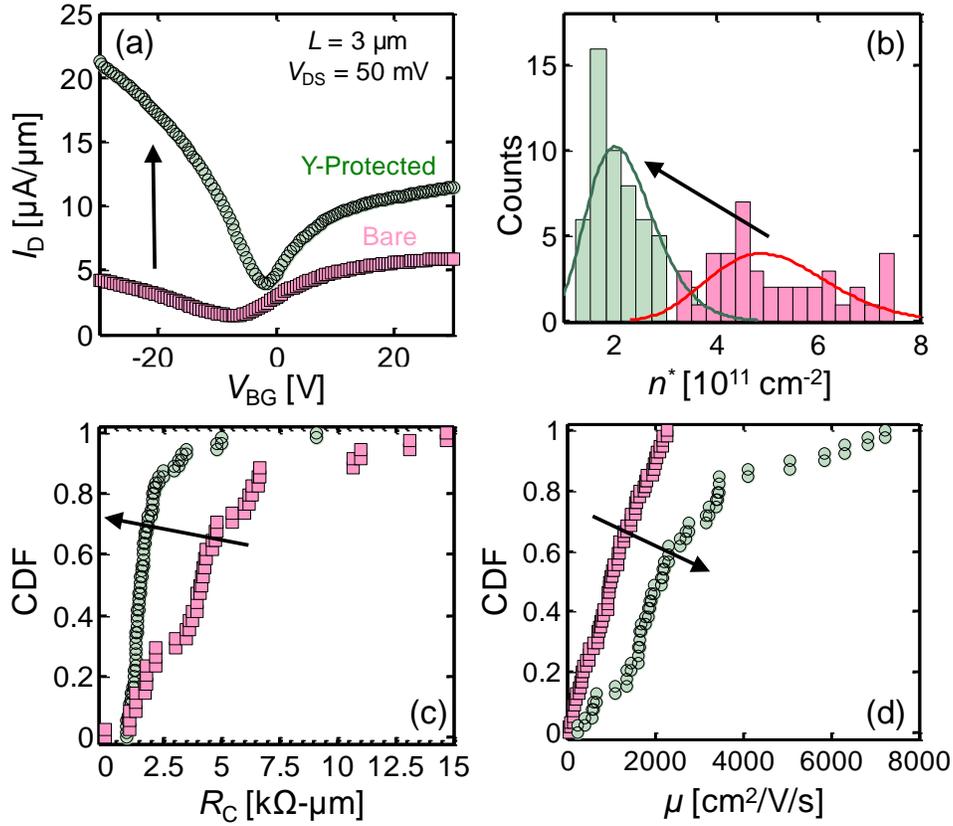

**FIG. 3.** Electrical data for 44 bare (pink squares) and 56 Y-protected (green circles) devices, with arrows denoting the change trend. (a) Measured current vs. back-gate voltage ($I_D$ vs. $V_{BG}$) for two devices displaying performance improvement of Y-protected devices. The Dirac voltage distributions of all bare and Y-protected devices are given in Supplementary Fig. S8. (b) Histogram of estimated carrier puddle density $n^*$, which is reduced in Y-protected devices to levels comparable to exfoliated graphene on $SiO_2$.[17,18] Cumulative distribution function (CDF) of *hole* (c) contact resistance and (d) mobility. The contact resistance, in particular, becomes smaller and more consistent for Y-protected devices. Distributions of *electron* contact resistance and mobility are shown in Supplementary Fig. S7.

**TABLE I.** Average electrical properties (hole mobility, contact resistance, carrier puddle density) and yield with two types of resists, Ti/Pd vs. pure Pd contacts, for bare and Y-protected devices. We note that $n^* = 2.2 \times 10^{11}$ cm$^{-2}$ corresponds to puddle surface potential variation $\Delta \approx \hbar v_F (\pi n^*)^{1/2} \approx 55$ meV, which is comparable to exfoliated graphene devices on SiO$_2$.[17,18]

| | LOL2000 Resist Lithography Ti/Pd Contacts | | | | PMGI Resist Lithography Pd Contacts | | | |
|---|---|---|---|---|---|---|---|---|
| | $\mu_h$ (cm$^2$V$^{-1}$s$^{-1}$) | $R_{C,h}$ (k$\Omega$-$\mu$m) | $n^*$ (cm$^{-2}$) | Yield | $\mu_h$ (cm$^2$V$^{-1}$s$^{-1}$) | $R_{C,h}$ (k$\Omega$-$\mu$m) | $n^*$ (cm$^{-2}$) | Yield |
| Bare | 1200 | 4.7 | $4.6 \times 10^{11}$ | 73% (44/60) | 1100 | 5.5 | $6.1 \times 10^{11}$ | 85% (55/64) |
| Y Protect. | 3100 | 1.9 | $2.2 \times 10^{11}$ | 97% (56/58) | 2600 | 2.2 | $4.3 \times 10^{11}$ | 93% (42/45) |

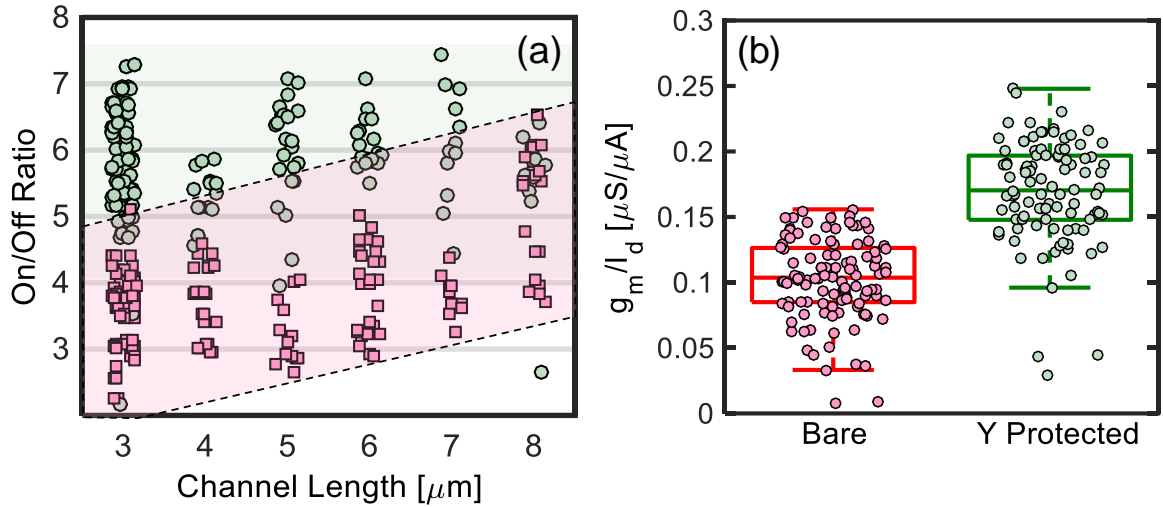

**FIG. 4.** (a) Measured on/off ratio of GFETs with various channel lengths with data laterally scattered for clarity. Pink squares represent bare devices and green circles represent Y-protected devices. The latter display up to 3× higher on/off ratio, independent of channel length due to lower $R_C$. (b) Y-sacrificial layer also increases analog GFET figure of merit, the max($g_m/I_D$) ratio. Low $g_m/I_D$ values result from thick $SiO_2$ oxide (90 nm) employed as gate dielectric, but the *relative* change due to process conditions is evident.

# Supplementary Material: Reducing Graphene Device Variability with Yttrium Sacrificial Layers

Ning C. Wang[1], Enrique A. Carrion[2], Maryann C. Tung[1], Eric Pop[1,3,a)]

[1]*Department of Electrical Engineering, Stanford University, Stanford, CA 94305, USA*

[2]*Department of Electrical and Computer Engineering & Micro and Nanotechnology Lab, University of Illinois at Urbana-Champaign, Urbana, IL 61801, USA*

[3]*Department of Materials Science & Engineering, Stanford University, Stanford, CA 94305, USA*

**Fabrication Details:** Monolayer graphene is first grown on Cu foil (99.8%, Alfa Aesar) via chemical vapor deposition (CVD) at 1000 °C in a 1" tube furnace. Two types of growths are conducted, at atmospheric pressure and at 100 mTorr conditions. Atmospheric growths use liquid isopropyl alcohol (IPA, reagent grade) in a metallic bubbler as a carbon source. The bubbler is chilled to 4 °C on a cold plate and 50 SCCM Ar serves as a carrier gas. 250 SCCM of Ar and 50 SCCM $H_2$ are mixed downstream, forming the necessary conditions for growth. Low pressure growths use a gaseous feedstock (100 SCCM $CH_4$) along with 50 SCCM $H_2$ and 1000 SCCM Ar. For both growths, the Cu foil is annealed in growth conditions without carbon sources for 1 hr.

Figure 1 in the main text summarizes the subsequent process steps. After growth, graphene is transferred onto 90 nm $SiO_2$ on $p^+$ Si ($< 0.005$ Ω·cm) by etching the Cu substrate in $FeCl_3$, using a PMMA scaffold for support and applying a modified RCA cleaning process[1] to minimize wrinkles and impurities (i.e. Fe, Cl, Cu). As a control in our study, a glass slide raised by silicon spacers shields half of each sample from Y evaporation, creating samples with no sacrificial layer [Fig. 1(b)].

We then fabricate back-gated graphene field-effect transistors (GFETs) using UV photolithography to define contact and channel regions, HCl to etch the sacrificial Y layer (where necessary), and e-beam evaporation for metal deposition of contacts. We use two different types of PR bilayer stacks (0.2/1.2 μm LOL2000/Shipley 3612 and 0.2/1.2 μm PMGI/Shipley 3612) in order to assess impact on devices electrical performance, as described in the main text.

After PR exposure/development and Y etching (samples are immersed in a dilute 10:1 deionized (DI) water to HCl solution for 15 s, although the photoresist survives baths up to 180 s without affecting feature definition), we deposit 1.5/40 nm Ti/Pd via e-beam evaporation (~$10^{-7}$ Torr) as contact metal. A separate set of devices using 40 nm Pd *without* a Ti interfacial layer was fabricated to study the impact of interfacial Ti adhesion layers on contact resistance. After 1 hour liftoff in Remover PG, we proceed with photolithography for channel definition (length $L = 3-10$ μm and width $W = 3-20$ μm). Finally, Y is removed in open regions after resist development to expose the underlying graphene to a subsequent $O_2$ plasma etch (100 SCCM $O_2$, 150 mTorr, 250 W) for graphene removal and channel definition. Finally, all Y is removed prior to device measurements with a prolonged 60 second dilute HCl bath.

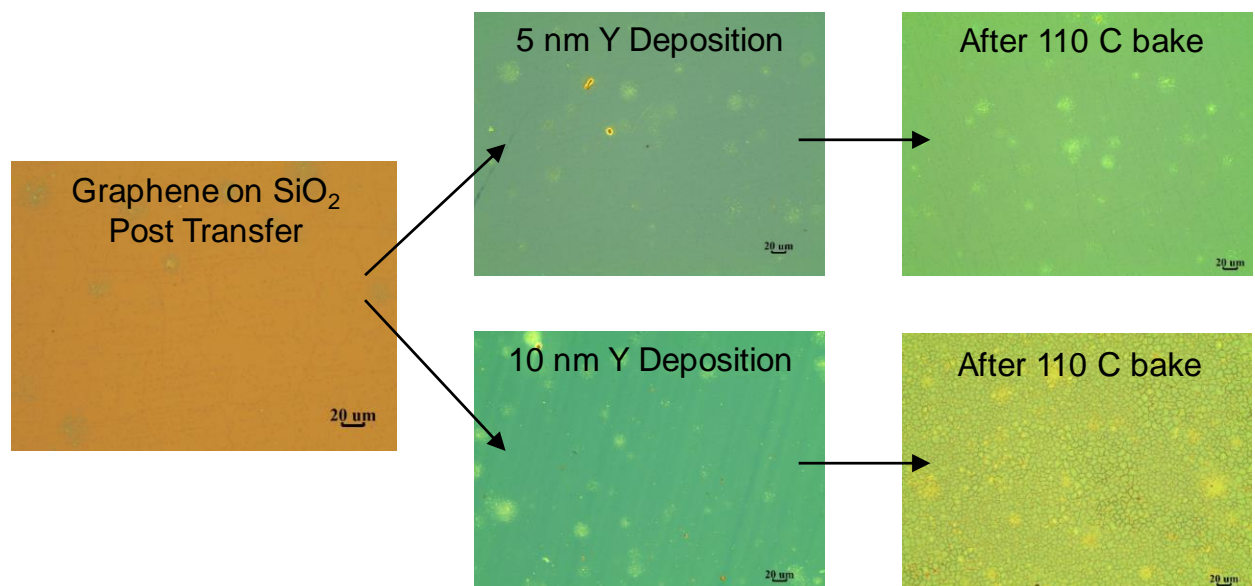

**Figure S2.** 20x optical images of initial graphene on SiO$_2$ sample, after 5 and 10 nm Y deposition, and after baking on a hot plate in air at 110 °C for 1 minute. Although residue spots are present on all samples, the spots are present on the initial graphene and are therefore not a result of the Y deposition. The hot plate step induces irreversible cracks in the 10 nm Y film, which can also affect the underlying graphene, most likely as a result of rapid oxidation and mismatches in coefficient of thermal expansion. However, the thinner ~5 nm Y film is unaffected, and thus we suggest keeping Y films < 5 nm to prevent cracking. Additional measures, such as baking in an inert environment, must be taken if a thicker film is necessary to prevent damage to graphene.

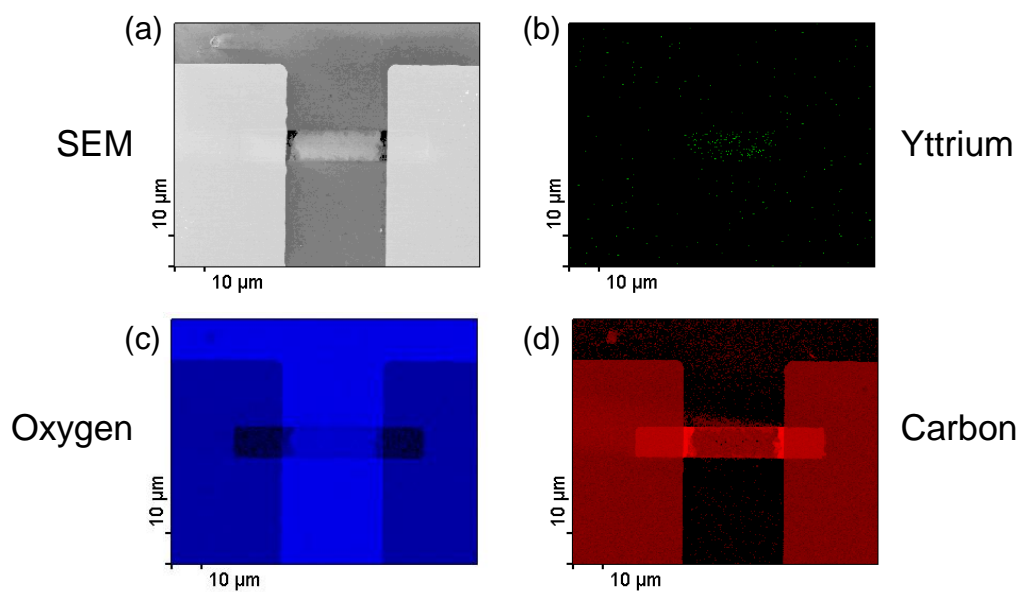

**Figure S3.** (a) SEM of device, followed by full Auger mapping of channel region after contact metallization and prior to final Y removal (Phi Instruments, 5 keV, 10 nA). (b) Yttrium mapping is averaged over 20 scans, while (c) oxygen and (d) carbon mappings are integrated over 5. Significant Y signal is present only in the channel in figure (b), confirming removal of Y in desired regions as well as remaining Y protecting the channel.

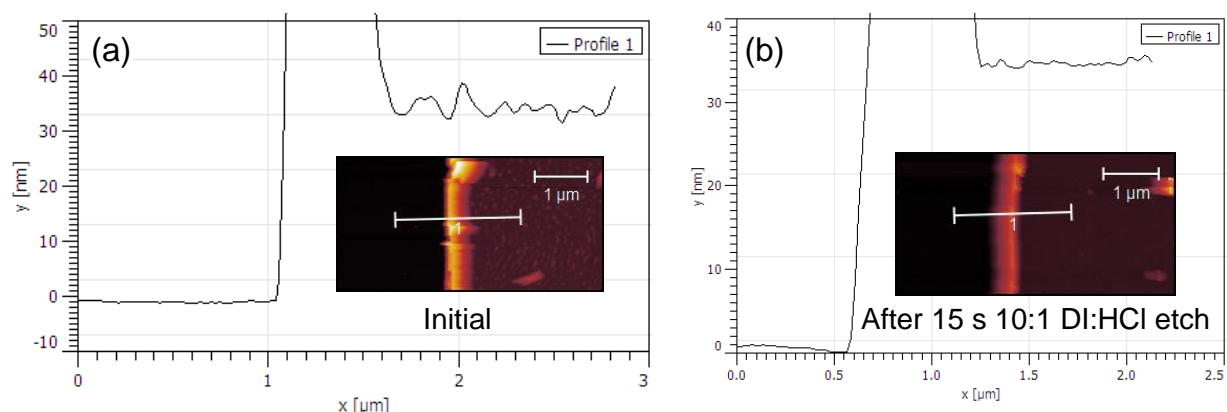

**Figure S4.** Tapping mode AFM (Digital Instruments) of Ti/Pd contacts (a) before and (b) after exposure to 10:1 DI:HCl for 15 seconds. Minimal Pd metal (~1 nm) is removed with the diluted solution, confirming a high etch selectivity (~ 3:1) to Y over surrounding contact metal.

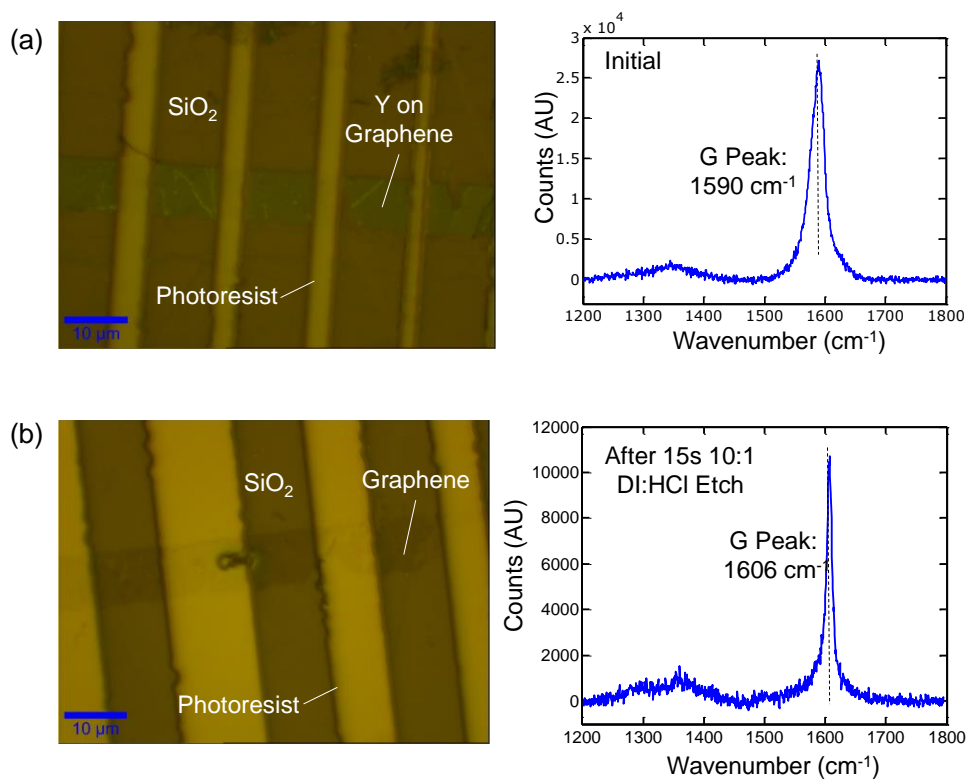

**Figure S5.** Raman spectroscopy (WiTec 532 nm laser) of graphene under defined contact region (a) before and (b) after Y removal in 10:1 DI:HCl. No noticeable increase in D-peak is observed after Y deposition, nor after HCl etch, confirming that the Y-sacrificial layer processing is safe for graphene. The G-peak is blue-shifted after exposure to HCl, indicative of some doping; however, this doping effect is reversible, as shown in Fig. S5.

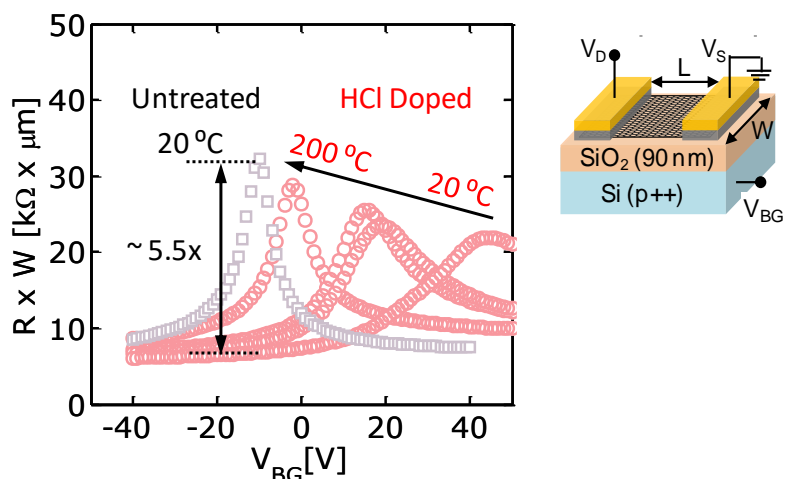

**Figure S6.** Electrical measurement of back-gated GFET before and after channel doping in 2:1 HCl for 1 second. Undoped device is shown as grey squares, while doped measurements are shown as pink circles. The doping effect is largely reversed by a 200 °C anneal in vacuum for 2 hours.

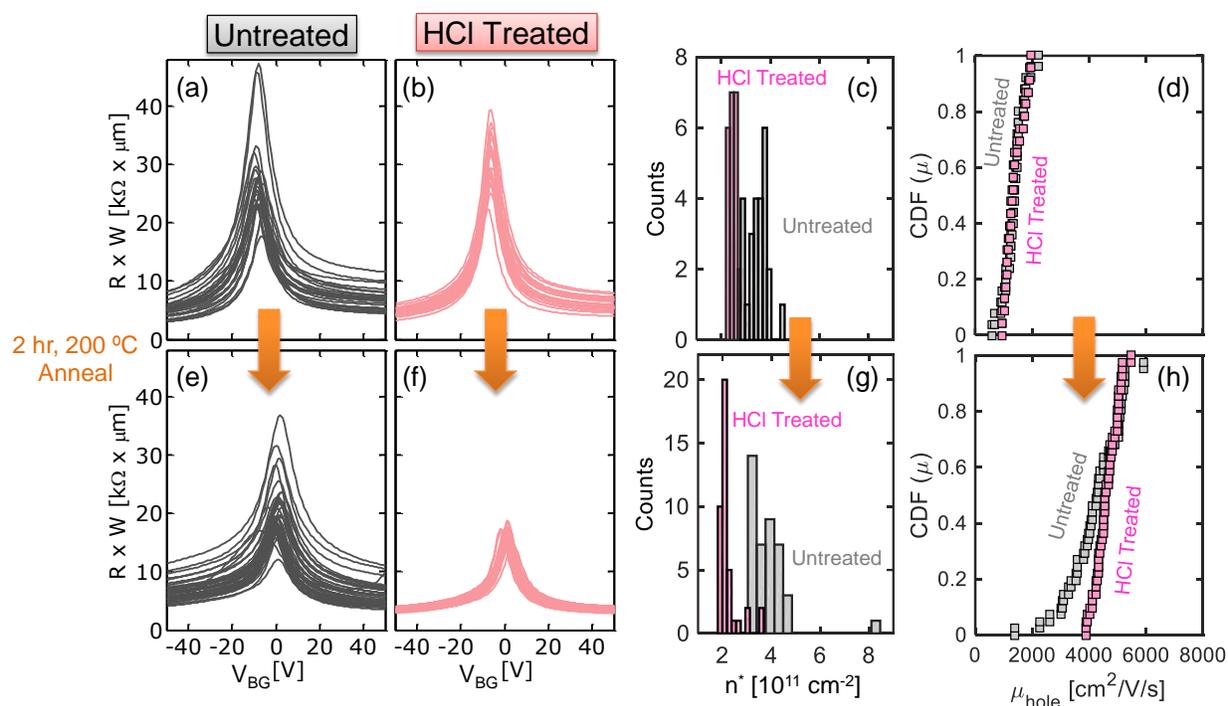

**Figure S6.** Electrical measurements (in ~$10^{-5}$ Torr vacuum) of back-gated GFET (a) with and (b) without 2:1 HCl treatment for 180 seconds. HCl treated devices exhibit (c) lower charge puddle concentration ($n^*$) than untreated devices, implying that the HCl treatment cleans the graphene surface, although (d) mobility is largely unaffected. (e) Untreated and (f) HCl treated devices exhibit slightly (g) lower charge puddle concentration with further vacuum annealing (2 hours, 200 °C). (h) Annealing improves hole transport similarly for untreated and HCl treated devices. Device channel length ($L$) is 2 μm and widths ($W$) range from 5 – 10 μm.

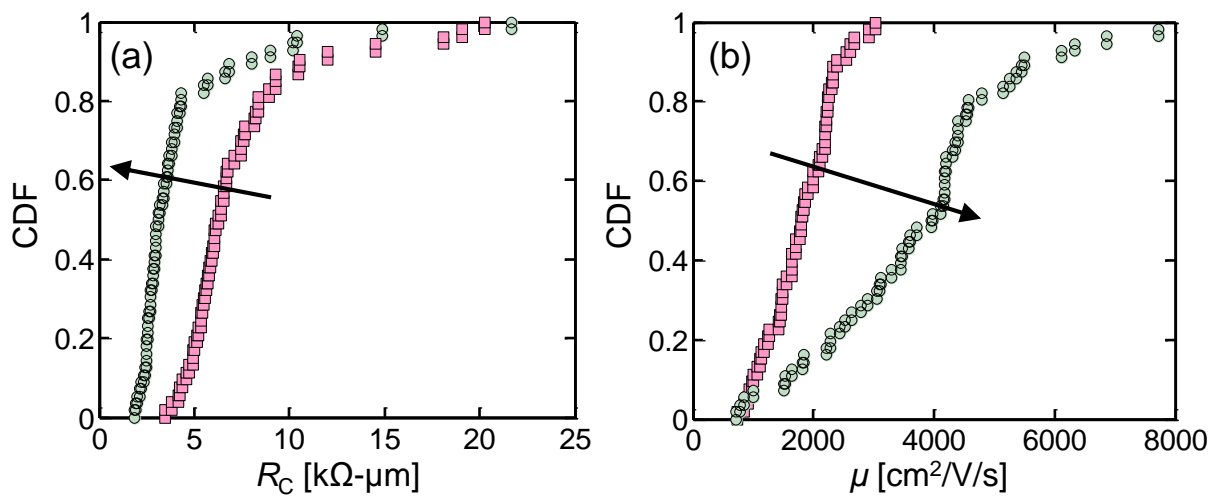

**Figure S7.** Cumulative distribution function (CDF) of *electron* contact resistance and mobility for devices with Ti/Pd contacts. 44 bare (pink squares) and 56 Y-protected (green circles) samples measured. Similar to hole transport (main text Fig. 3), electron transport properties improve as well with use of Y-sacrificial layers as (a) average $R_C$ decreases from 7.4 to 4.3 kΩ-μm and (b) average $\mu$ increases from 1792 to 3730 cm$^2$V$^{-1}$s$^{-1}$.

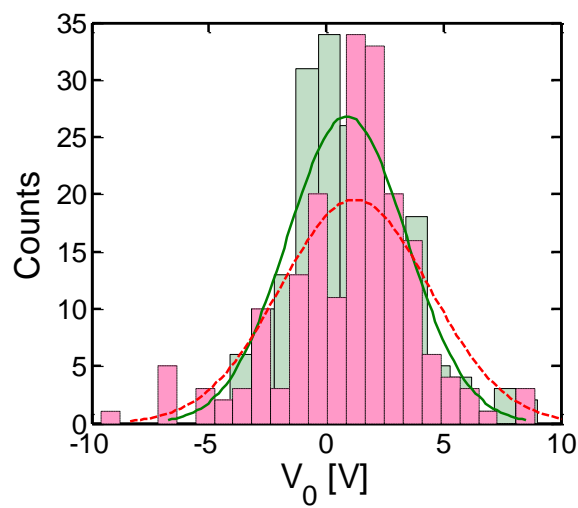

**Figure S8.** Histogram of charge neutrality point (CNP) i.e. Dirac voltage $V_0$ for all 378 devices measured in this study. The dashed red line for bare devices and the solid green line for Y protected devices represent fitted Gaussian distributions. The average $V_0$ is reduced from 1.22 to 0.89 V and the standard deviation decreased from 3.19 to 2.54 V with the Y-sacrificial process. These values are measured on 90 nm SiO$_2$ back-gate oxide, thus values for ~1 nm equivalent oxide thickness (EOT) would be reduced by a factor of 90x (see main text).

**Table S1.** Average values of electrical data (electrons, holes, mobility, contact resistance, puddle charge density, CNP voltage) for all 378 GFETs fabricated from various graphene sources.

| | $\boldsymbol{\mu}_e$ [cm$^2$V$^{-1}$s$^{-1}$] | $\boldsymbol{M}_h$ [cm$^2$V$^{-1}$s$^{-1}$] | $\boldsymbol{R}_{C,e}$ [kΩ-µm] | $\boldsymbol{R}_{C,h}$ [kΩ-µm] | $n^*$ [cm$^{-2}$] | $V_0$ [V] |
|---|---|---|---|---|---|---|
| **Stanford University (Atmospheric Pressure CVD)** | | | | | | |
| Bare | 1800 | 1200 | 7.4 | 4.7 | $4.6 \times 10^{11}$ | -1.4 |
| **Y-Sacr.** | **3700** | **3100** | **4.3** | **1.9** | **$2.2 \times 10^{11}$** | **-0.8** |
| **University of Illinois Urbana-Champaign (Low Pressure ~100 mTorr CVD)** | | | | | | |
| Bare | 2100 | 1270 | 10.2 | 6.1 | $5.0 \times 10^{11}$ | 2.5 |
| **Y-Sacr.** | **3000** | **2700** | **6.4** | **3.5** | **$2.8 \times 10^{11}$** | **0.6** |
| **Graphene Supermarket** | | | | | | |
| Bare | ---* | ---* | 17.9 | 4.3 | $7.3 \times 10^{11}$ | 13.5 |
| **Y-Sacr.** | **3700** | **2300** | **4.3** | **1.3** | **$4.6 \times 10^{11}$** | **6.1** |

*Data not shown due to poor fit.